\documentclass[12pt]{elsarticle}
\makeatletter
\def\ps@pprintTitle{%
 \let\@oddhead\@empty
 \let\@evenhead\@empty
 \def\@oddfoot{}%
 \let\@evenfoot\@oddfoot}
\makeatother
\usepackage{graphicx}
\usepackage{mathptmx}
\usepackage[intlimits,centertags]{amsmath}
\usepackage{amssymb,amsfonts}
\usepackage[pdftex]{hyperref}
\usepackage{aas_macros}
\usepackage{enumerate}
\usepackage{xfrac}
\usepackage[table,xcdraw]{xcolor}
\usepackage{sectsty}
\usepackage{subcaption}
\usepackage[normalem]{ulem}
\usepackage{setspace}
\usepackage[verbose]{wrapfig}
\usepackage{lipsum}
\usepackage{ragged2e}
\usepackage{etoolbox} 
\usepackage{combelow}
\usepackage{lineno}
\usepackage[symbol]{footmisc}
\usepackage[T1]{fontenc}
\usepackage{pdfpages}
\usepackage{times}
\usepackage[compact]{titlesec}
\usepackage{geometry}
\usepackage{fdsymbol}
\usepackage[utf8]{inputenc}
\usepackage{enumitem,amssymb}
\usepackage{ragged2e}
\usepackage[symbol]{footmisc}

\newlist{thematic}{itemize}{8}
\setlist[thematic]{label=$\square$}
\usepackage{pifont}

\newif\ifastrophysical
\astrophysicalfalse

\newif\iffundamental
\fundamentaltrue

\biboptions{sort&compress}

\definecolor{header_color}{HTML}{15588c}
\subsectionfont{\color{header_color}}
\renewcommand\refname{\color{header_color}References}

\begin{document}

\title{Neutrino astronomy with the next generation IceCube Neutrino Observatory}

\begin{abstract}
The past decade has welcomed the emergence of cosmic neutrinos as a new messenger to explore the most extreme environments of the universe. The discovery measurement of cosmic neutrinos, announced by IceCube in 2013, has opened a new window of observation that has already resulted in new fundamental information that holds the potential to answer key questions associated with the high-energy universe, including: what are the sources in the PeV sky and how do they drive particle acceleration; where are cosmic rays of extreme energies produced, and on which paths do they propagate through the universe; and are there signatures of new physics at TeV-PeV energies and above? The planned advancements in neutrino telescope arrays in the next decade, in conjunction with continued progress in broad multimessenger astrophysics, promise to elevate the cosmic neutrino field from the discovery to the precision era and to a survey of the sources in the neutrino sky. The planned detector upgrades to the IceCube Neutrino Observatory, culminating in IceCube-Gen2 --- an envisaged \$400M facility with anticipated operation in the next decade, described in this white paper --- are the cornerstone that will drive the evolution of neutrino astrophysics measurements.
\end{abstract}


\author[christchurch]{M. G. Aartsen\fnref{}}
\author[zeuthen]{M. Ackermann\fnref{}}
\author[christchurch]{J. Adams\fnref{}}
\author[brusselslibre]{J. A. Aguilar\fnref{}}
\author[copenhagen]{M. Ahlers\fnref{}}
\author[stockholmokc]{M. Ahrens\fnref{}}
\author[geneva]{C. Alispach\fnref{}}
\author[marquette]{K. Andeen\fnref{}}
\author[pennphys]{T. Anderson\fnref{}}
\author[brusselslibre]{I. Ansseau\fnref{}}

% ...
\address[brusselslibre]{Universit{\'e} Libre de Bruxelles, Science Faculty CP230, B-1050 Brussels, Belgium}
% ...
\address[christchurch]{Dept. of Physics and Astronomy, University of Canterbury, Private Bag 4800, Christchurch, New Zealand}
% ...
\address[copenhagen]{Niels Bohr Institute, University of Copenhagen, DK-2100 Copenhagen, Denmark}
% ...
\address[geneva]{D{\'e}partement de physique nucl{\'e}aire et corpusculaire, Universit{\'e} de Gen{\`e}ve, CH-1211 Gen{\`e}ve, Switzerland}
% ...
\address[marquette]{Department of Physics, Marquette University, Milwaukee, WI, 53201, USA}
% ...
\address[stockholmokc]{Oskar Klein Centre and Dept. of Physics, Stockholm University, SE-10691 Stockholm, Sweden}
% ...
\address[pennphys]{Dept. of Physics, Pennsylvania State University, University Park, PA 16802, USA}
% ...
\address[zeuthen]{DESY, D-15738 Zeuthen, Germany}
\fntext[padova]{also at Universit{\`a} di Padova, I-35131 Padova, Italy}
\fntext[mephi]{also at National Research Nuclear University, Moscow Engineering Physics Institute (MEPhI), Moscow 115409, Russia}

\pagenumbering{roman}
\maketitle
\normalsize
\pagebreak

\pagenumbering{arabic}
\setcounter{footnote}{0}
With the first detection of high-energy neutrinos of extraterrestrial origin in 2013~\cite{Aartsen:2013jdh}, the IceCube Neutrino Observatory opened a new window to some of the most extreme phenomena of our universe. Neutrinos interact only weakly with matter and therefore escape energetic and dense astrophysical environments that are opaque to electromagnetic radiation. In addition, at PeV (10$^{15}$~eV) energies, extragalactic space becomes opaque to electromagnetic radiation due to the scattering of high-energy photons ($\gamma$-rays) on the cosmic microwave background and other radiation fields. This leaves neutrinos as the only messengers available to search for the most extreme particle accelerators in the cosmos --- the sources of the ultra high energy cosmic rays (CRs). These CRs reach energies of more than $10^{20}$~eV, which is a factor of $10^{7}$ times higher than at the most powerful man-made particle accelerators. 

\begin{wrapfigure}[24]{r}{7cm}
\centering\includegraphics[width=\linewidth]{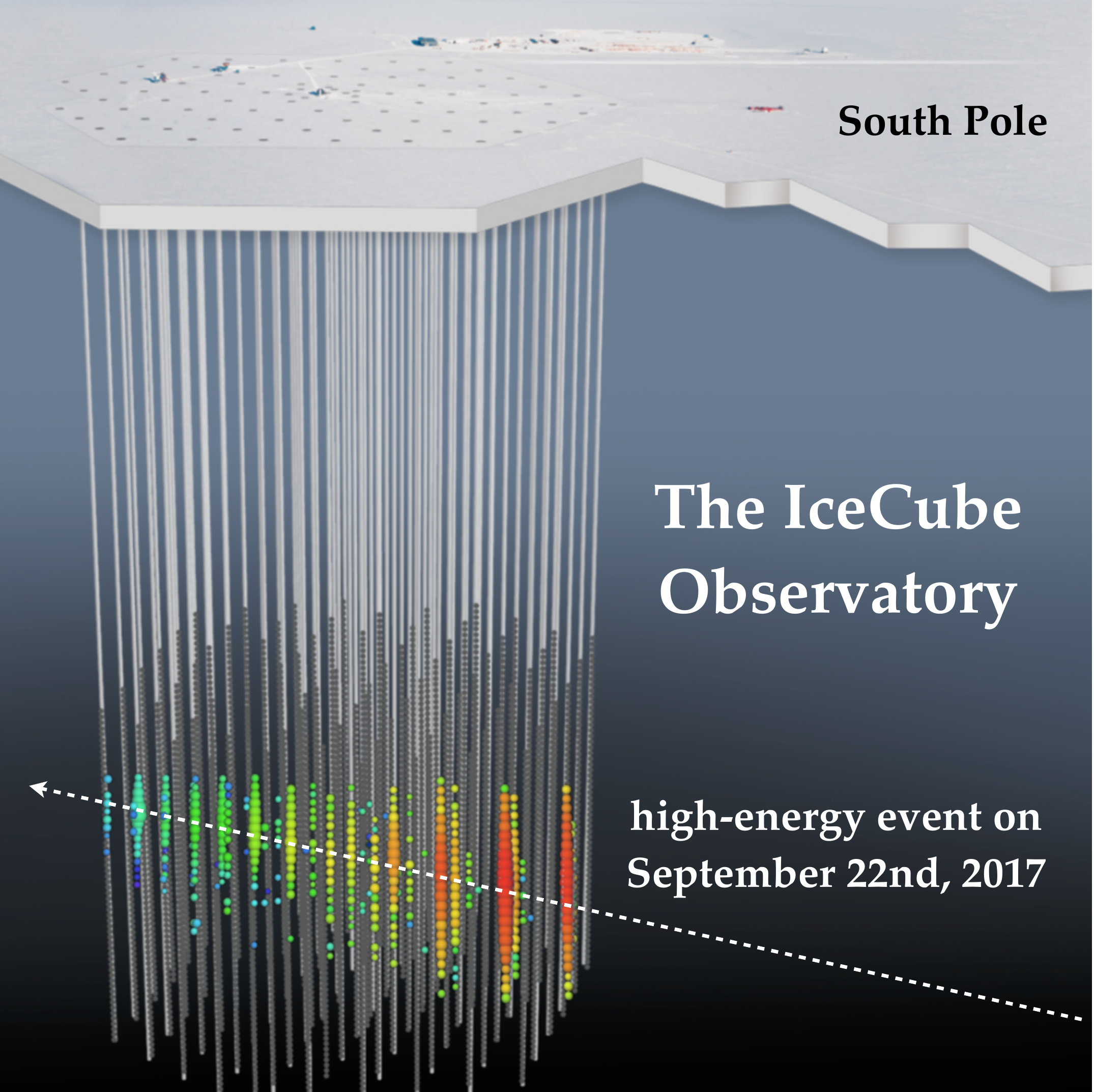}
\caption{Schematic illustration of the current IceCube array, located near the geographic
South Pole at depths of up to 2.5\,km. The event view of the high-energy cosmic neutrino that triggered multimessenger observations of TXS~0506+056 is also shown, where the colored spheres each represent the timing---red is early, blue is late---of an optical module that has observed light from the muon produced in the neutrino's interaction somewhere outside the detector.}
\label{ic170922a}
\end{wrapfigure}

\begin{figure}[hbt!]
\centering\includegraphics[width=15cm]{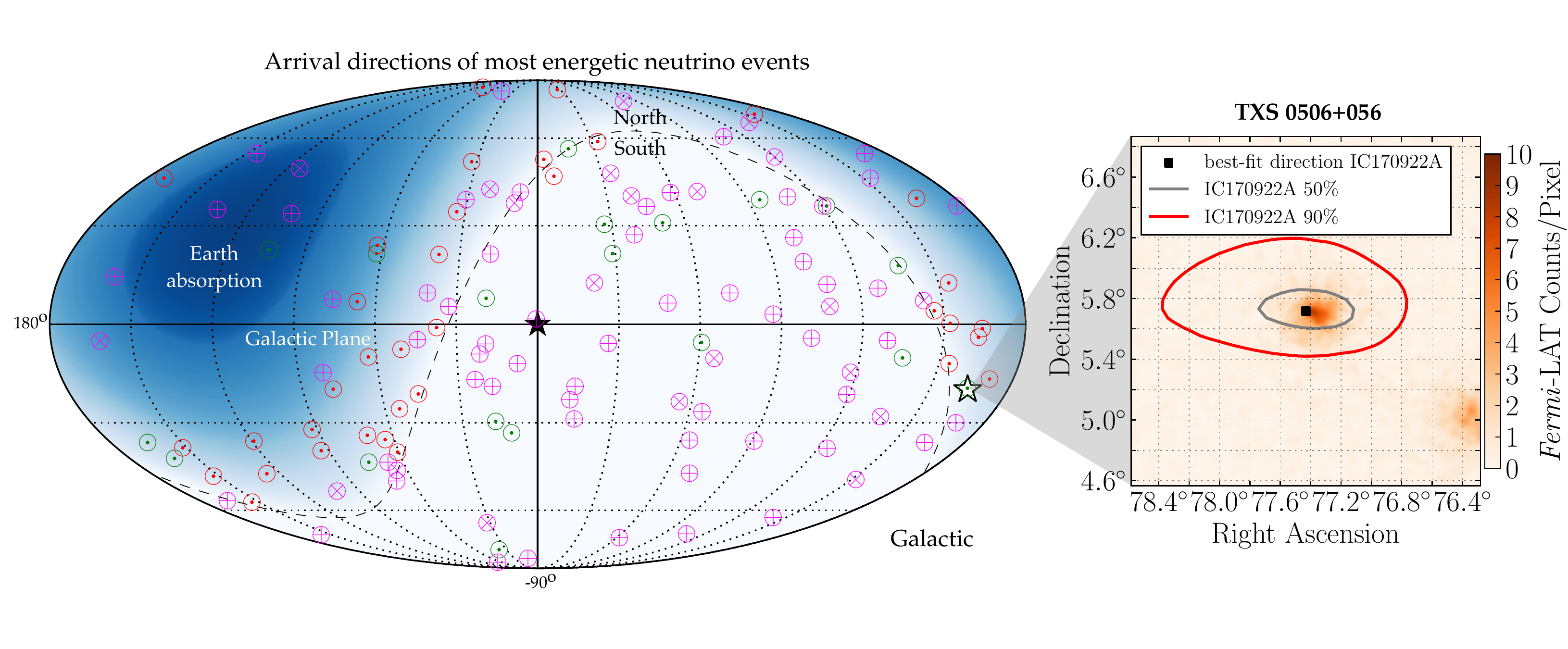}
\caption{The current sky map of highly energetic neutrino events detected by IceCube. The distribution of the events is consistent with being isotropic. The location of the first compelling neutrino source, blazar TXS~0506+056, is marked with a star.  Shown in the inset are the related $Fermi$-LAT measurements of the region centred on TXS~0506+056 from September 2017~\cite{IceCube:2018dnn}.  The uncertainty ellipses of the IceCube neutrino event IC-170922A are shown for reference.}
\label{skymap_fermiexplosion}
\end{figure}

High energy neutrinos are produced through the interaction of CRs in the sources with ambient matter or radiation fields. Unlike the charged CRs, neutrinos are not deflected by magnetic fields on the way from the source to the Earth, but point back to their origin, thus providing for a smoking-gun signature of CR acceleration. The power of this approach was recently demonstrated by IceCube and the broad astronomical community when a single high-energy neutrino (see Fig.\,\ref{ic170922a})
was observed in coincidence with a flaring GeV-blazar, revealing what appears to be the first known extragalactic source of high energy cosmic rays (see Fig.~\ref{skymap_fermiexplosion})~\cite{IceCube:2018dnn, IceCube:2018cha}. 

In order to evolve beyond the initial discovery era of astronomy with neutrinos, it is necessary to improve the sensitivity of the detector in a few of the key areas that neutrino telescopes share with the broad suite of astronomical instruments: exposure and pointing resolution. For the current IceCube detector, providing full-sky (dependent on energy) coverage with better than 99.5\% up-time, a few tens of well-reconstructed high-energy astrophysical events\,
are observed and issued as near real-time alerts each year with angular resolutions at or below 1 degree. Compared to typical astronomical instruments,
this angular resolution introduces a large uncertainty when identifying single sources, in particular in heavily populated regions of the sky. The IceCube Upgrade, fully funded and currently entering the construction phase, provides the first important step in closing this gap via an advanced calibration program. 

The science and the resulting requirements for a next generation neutrino observatory were highlighted in several science White Paper contributions submitted to the Astro2020 Decadal Survey by members of the astronomical community. These contributions pursue a diverse set of research topics, ranging from a focus on neutrino astronomy~\cite{2019BAAS...51c.185V} and fundamental physics with cosmic neutrinos~\cite{2019BAAS...51c.215V}, to those on cosmic ray science~\cite{2019BAAS...51c.131S, 2019astro2020T.459F, 2019BAAS...51c..93S}, those providing a comprehensive view of extragalactic ~\cite{2019BAAS...51c..92R,2019BAAS...51c.228S,2019BAAS...51c.396V} as well as Galactic sources~\cite{2019BAAS...51c.115C,2019BAAS...51c.194N}, and focusing on multimessenger studies of sources with $\gamma$-rays~\cite{2019BAAS...51c.553V,2019BAAS...51c.267H,2019BAAS...51c.431O,2019astro2020T.427C} or gravitational waves~\cite{2019BAAS...51c..38B, 2019BAAS...51c.239K, 2019BAAS...51c.247F}. 

\begin{wrapfigure}{R}{7.5cm}
\centering
\includegraphics[width=\linewidth]{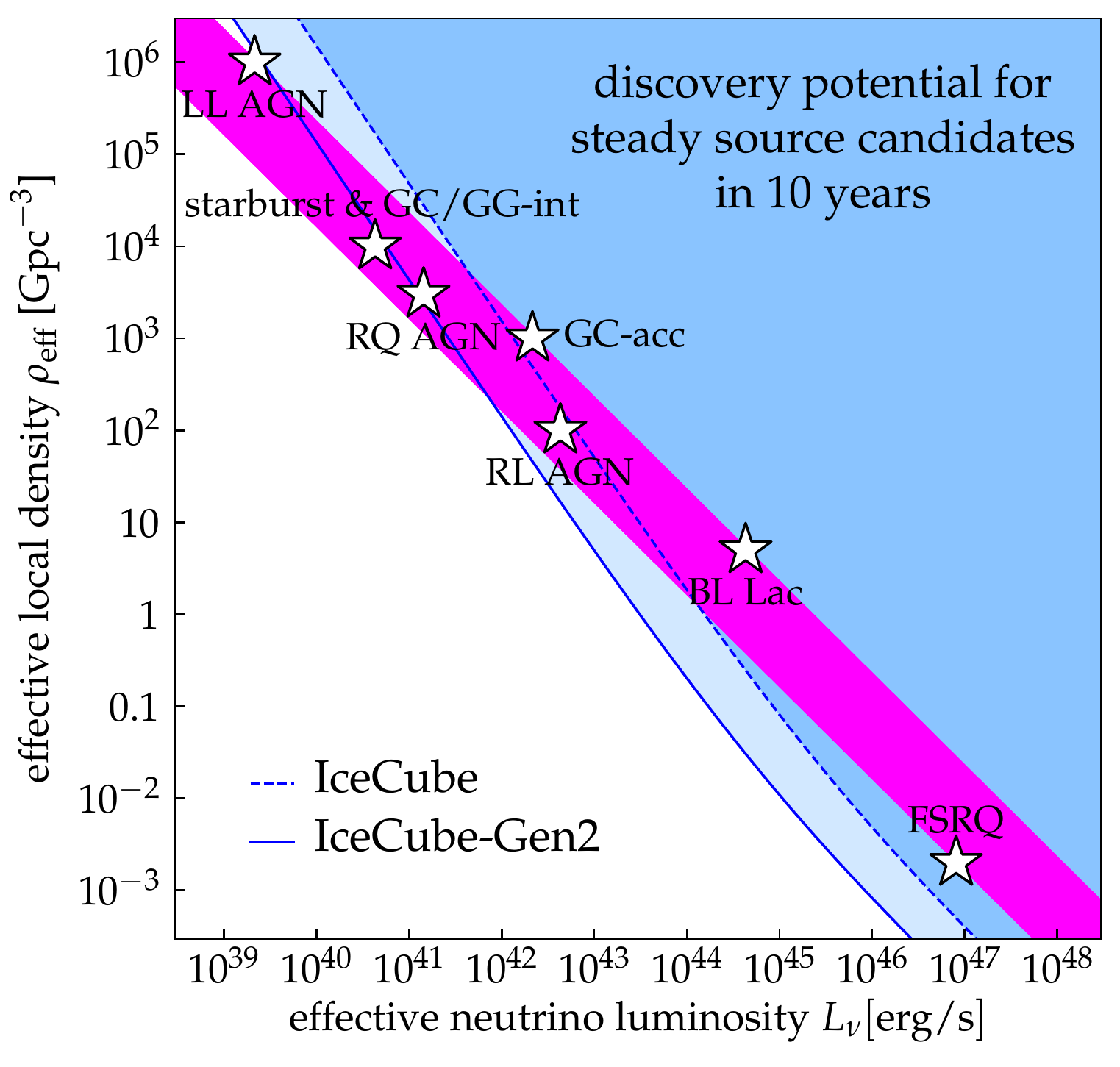}
\caption{Comparison of the diffuse neutrino emission (solid magenta band) to the effective local density and luminosity of candidate extragalactic neutrino source populations. We indicate several populations ($\medwhitestar$) by the required neutrino luminosity to account for the full diffuse flux~\cite{Murase:2016gly} (see also \cite{Silvestri:2009xb}). The lower (upper) edge of the band assumes rapid (no) redshift evolution. The shaded regions indicate IceCube's (dark-blue, dashed line) and IceCube-Gen2's (light-blue, solid line) ability to discover one or more  of the population ($E^2\phi_{\nu_\mu+\bar\nu_\mu}\simeq 10^{-12}~{\rm TeV}/{\rm cm}^2/{\rm s}$ in the Northern Hemisphere~\cite{Aartsen:2018ywr}).}
\label{astro_constraints}
\end{wrapfigure}

To push to the ultimate goal of everyday neutrino-based astronomy will require collecting a factor of 5 to 10 more neutrinos than an IceCube-scale detector, with a concommitant sensitivity increase to point sources of at least 5 times the current IceCube array (see Fig.~\ref{astro_constraints}) as well as an expansion of the energy range to beyond $10^{18}$\,eV energies with sensitivity two order of magnitude better than what is currently accessible (see  \cite{2019BAAS...51c.185V} for details).

IceCube-Gen2, a planned next generation observatory at the South Pole, will achieve these goals. Construction of the extended observatory is estimated to require 6 years, with completion and full operation at the end of the next decade, and with a total project budget of approximately \$400M. IceCube-Gen2 is anticipated to play an essential role in shaping the new era of multimessenger astronomy in that period to resolve a number of the most pressing questions of the high-energy universe ($\it{e.g.}$, identify the sources  of the mostly unresolved IceCube flux, see Fig.\,\ref{astro_constraints}). This project mission relies on operating in concert with the anticipated new survey instruments across the electromagnetic spectrum and 
gravitational wave instruments becoming available in the next decade.

\subsection*{Current era: The IceCube Neutrino Observatory and High-Energy Cosmic Neutrinos}
Constructed via an NSF MREFC grant between 2004 and 2010, IceCube instruments
one cubic kilometer of the deep glacial ice near the Amundsen-Scott South Pole Station, Antarctica. A total of 5160 digital optical modules (DOMs), each autonomously operating a 25 cm photomultiplier tube (PMT) in a glass pressure housing~\cite{Aartsen:2016nxy}, are currently deployed at depths between 1450\,m and 2450\,m on 86 cables (‘strings’). The glacial ice behaves as both the interaction medium and support structure for the IceCube array. Cherenkov radiation emitted by secondary charged particles produced in neutrino interactions in or near the active detector volume carries the information of the neutrino’s energy, direction, arrival time, and flavor. Digitized waveforms from each DOM provide the record of the event signatures in IceCube, including the arrival time and amplitude (charge) of the detected Cherenkov photons emitted by the charged particles as they traverse the ice.

With a cubic-kilometer instrumented volume, IceCube is more than an order of magnitude larger than previous and current experiments operating in the North (Baikal Deep Under-water Neutrino Telescope~\cite{Belolaptikov:1997ry}, ANTARES~\cite{Collaboration:2011nsa}) and at the South Pole (AMANDA-II~\cite{Andres:2001ty}), and of similar size as current projects under construction (KM3NeT~\cite{Adrian-Martinez:2016fdl}, GVD~\cite{Avrorin:2013uyc}).
It has collected neutrino-induced events with energies of a few GeV to beyond 10 PeV, the latter
corresponding to the highest-energy neutrinos ever observed and opening new scientific avenues not just for astronomy but also for probing physics beyond the standard model of particle physics (see, {\it e.g.}, \cite{Aartsen:2017kpd}). 
Independent evidence for astrophysical neutrinos comes from different detection channels, including shower-like events~\cite{Niederhausen:2017mjk}, events that start inside the instrumented volume~\cite{Aartsen:2014gkd}, through-going events~\cite{Aartsen:2015rwa}, as well as first ``double-bang'' tau-neutrino events \cite{Stachurska:2019srh} that are not expected to be produced in the atmosphere through conventional channels. While the collective significance for the cosmic origin of the neutrinos is now beyond doubt, a decade of IceCube data taking underscores the rarity of the measurements; {\it e.g.}, only one tau neutrino and one electron anti-neutrino at the Glashow resonance~\cite{Glashow:1960zz} have been observed to date. Clearly, much larger statistics are needed to exploit the full potential of all-flavor neutrino astronomy.     

The low rate of cosmic neutrinos and the moderate angular resolution of $\sim0.5^{\circ}$ for muon neutrinos and $\sim 10^{\circ}$ for electron and tau neutrinos (so called cascade-like events) make identification of neutrino point sources challenging.
Although an essential clue has been extracted from the data with respect to the blazar TXS~0506+056 (see below), the spatial distribution of astrophysical neutrinos detected by IceCube is largely consistent with isotropy (see Fig.~\ref{ic170922a}),
implying that a substantial fraction of IceCube’s cosmic neutrinos are of extragalactic origin.

The most compelling evidence for a neutrino point source to date is the detection of one neutrino event (IC-170922A) in spatial and temporal coincidence with an enhanced $\gamma$-ray emission state of the blazar TXS~0506+056 \cite{IceCube:2018dnn}. Evidence for another period of enhanced neutrino emission from this source, in 2014/15, was revealed in a dedicated search in the IceCube archival data \cite{IceCube:2018cha}. The individual chance probabilities of the blazar-neutrino association and the observed excess in the IceCube data alone are each at a significance level of 3-3.5$\sigma$.

Further observations of a similar nature are necessary to provide definitive statements about the production mechanism of neutrinos in blazars. It is becoming increasingly clear that the neutrino signal from blazars is not a predominant component of current IceCube data, otherwise the existing correlation studies should have already provided indications of a more significant signal \cite{IceCube:2018dnn, Aartsen:2019gxs, Padovani:2016wwn}.
As an example, a comparison of the full set of IceCube neutrinos with a catalog of $\gamma$-ray blazars does not produce evidence of a correlation and results in an upper bound of $\sim$30\% as the maximum contribution from these blazars to the diffuse astrophysical neutrino flux below 100 TeV~\cite{Glusenkamp:2015jca}. Accordingly, a blazar population responsible for the neutrinos would have to be appropriately dim in $\gamma$-rays (see {\it e.g.}~\cite{Halzen:2018iak, Neronov:2018wuo}). 
Another widely considered candidate source of extragalactic neutrinos are $\gamma$-ray bursts (GRBs). Similar to blazars, the non-detection of neutrinos in spatial and temporal coincidence with GRBs over many years has placed a strict upper bound of 1\% for the maximum contribution from observed GRBs to the diffuse flux observed by IceCube~\cite{Aartsen:2014aqy}. 

Other source populations are therefore anticipated to contribute to the dominant fraction of the astrophysical neutrino flux. Indirect constraints may be derived for $\gamma$-ray transparent sources. Since $\gamma$-rays and neutrinos are produced in the same hadronic interaction, one can expect that, if the $\gamma$-rays can escape from the source, the bolometric luminosity in $\gamma$-rays is at least as large as in neutrinos. Over cosmological distances, high-energy $\gamma$-rays will cascade to lower energies in the intergalactic radiation fields, and eventually reach Earth in the energy band above 10 GeV. The $Fermi$-LAT measurement of the total extragalactic $\gamma$-ray emission \cite{Ackermann:2014usa} in this band therefore also constrains the flux of neutrinos from $\gamma$-ray transparent sources. Careful modeling indicates that the measured spectrum of astrophysical neutrinos could be associated with an overproduction of $\gamma$-rays unless the sources  are dark in $\gamma$-rays, {\it i.e.}, so-called hidden sources~\cite{Murase:2015xka}. This connection to $\gamma$-rays constrains the contribution of $\gamma$-transparent sources such as starburst galaxies or galaxy clusters, making them less likely the main source of IceCube's extraterrestrial neutrinos.  

It is clear that the nature of the neutrino sky is complex, with the mystery of the sources and production mechanism of the high-energy neutrino flux largely unresolved. In the following we describe the proposed upgrade of IceCube to a scale and performance level needed to resolve these key unknowns in modern astrophysics.
 
\subsection*{Near Future: The IceCube Upgrade (IceCube-Gen2 Phase 1)}
\begin{figure}[htb!]
\centering\includegraphics[width=0.8\linewidth]{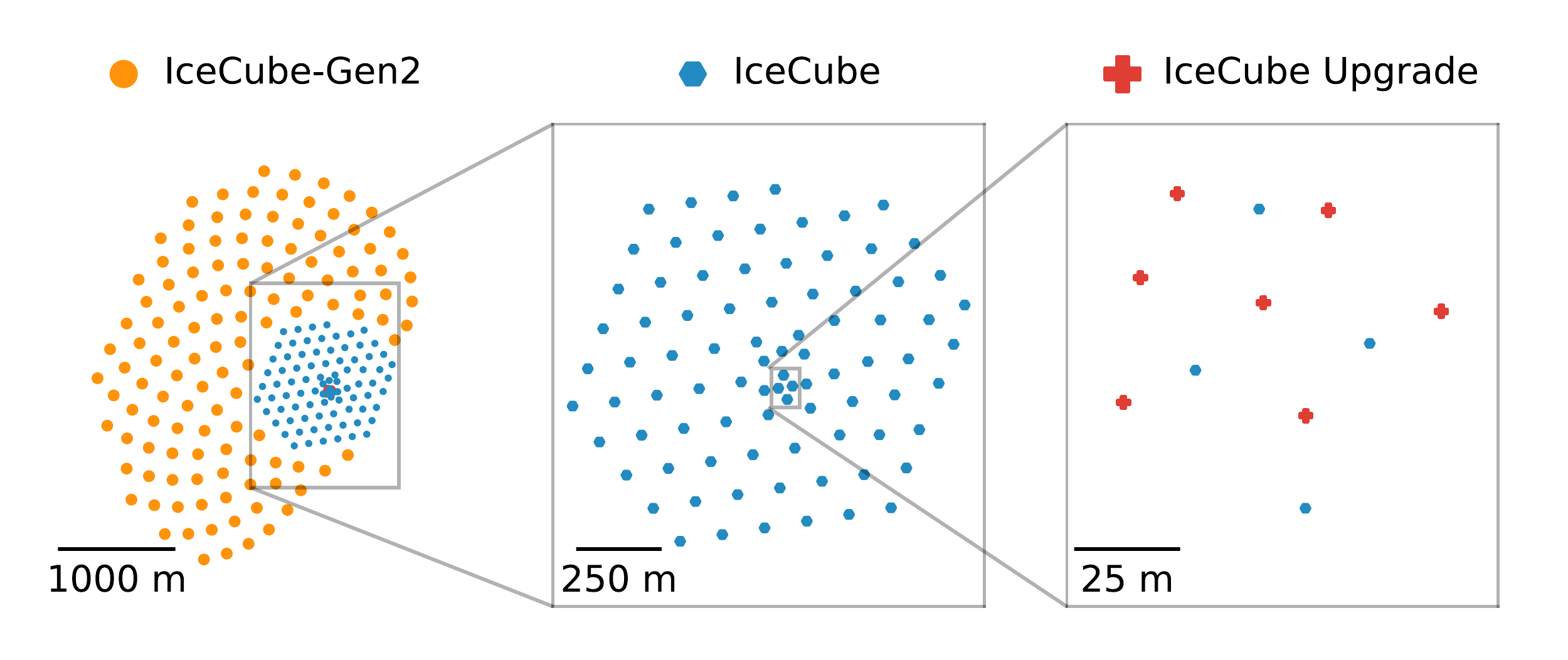}
\caption{Top view of the IceCube-Gen2 optical in-ice arrays. \textbf{Left}: The IceCube-Gen2 strings in the high-energy array “sunflower” layout (orange). In the initial design studies, 120 new strings are spaced 240 m apart and nominally instrumented with 80 IceCube-type optical modules over a vertical length of 1.25 km. The total instrumented volume in this design is 7.9 km$^3$, approaching an order of magnitude larger than IceCube alone.
\textbf{Center}: The current IceCube detector array (blue), including the DeepCore array with reduced spacing in the center. Above each IceCube string location are two IceTop tanks that operate together as a large-scale cosmic-ray air shower detector and partial atmospheric muon veto for the in-ice array. 
\textbf{Right}: The seven IceCube Upgrade strings (red) relative to existing IceCube and DeepCore strings. }
\label{ic_detector}
\end{figure}

The IceCube Upgrade (see Fig.\,\ref{ic_detector}), currently entering the construction phase, utilizes seven new strings of advanced instrumentation. 
The seven Upgrade strings are designed to be densely instrumented near the bottom-center of the existing IceCube array to provide improved detection of unscattered photons from particle interactions above approximately 1$~$GeV. The improved detection efficiency will provide world-leading sensitivity to measurements of neutrino properties via oscillations. In particular, the projected Upgrade measurements of tau neutrino appearance via atmospheric neutrino oscillations will provide stringent tests of the unitarity of the  neutrino mixing parameters, complementing the anticipated measurements from the global long-baseline neutrino community during the same period over the next decade. For astrophysical neutrinos, the Upgrade's scientific program also includes as a primary element a robust calibration program that is designed to improve the knowledge of the natural ice medium, making it possible to enhance reconstruction algorithms and control current leading systematic uncertainties. The ability to  re-calibrate the existing and future IceCube data is expected to enhance typical angular resolutions for high-energy neutrinos of 
better than $0.3^{\circ}$ for track-like events and better than $5^{\circ}$ for cascade-like events. 
Given this, as the pathfinder towards a full Gen2 array, the IceCube Upgrade will employ technology that will simultaneously augment the on-going (and long-term measurements) of the general IceCube facility while also providing the benefits of an ideal test environment for the R\&D efforts needed to realize IceCube-Gen2.

\subsection*{Decadal Vision: The IceCube-Gen2 Observatory}

A second generation observatory, IceCube-Gen2, addresses the two main limiting factors in the current IceCube instrument: event rate and angular resolution. It does so by improving the event rate by a factor of 4-10, depending on channel, and angular resolution by a factor of approximately three. Together, this results in sensitivity to sources five times fainter than visible today.
Further, through the addition of an ultra-high-energy radio array, IceCube-Gen2 will expand the accessible energy range of cosmic neutrinos by several orders of magnitude compared to IceCube.

\noindent\textbf{In-ice optical array (OA):} The basic concept for the Gen2 design starts with the idea that the primary science goals will target higher energies (from approximately 10\,TeV to 10\,EeV) than those for which IceCube was designed, 
while, through advanced calibration techniques, continuing to enhance
the knowledge of the the natural ice medium and of the detector response to provide a direct evolution of the near-real-time multimessenger alert system for the low-background high-energy regime. IceCube has demonstrated that the diffuse cosmic neutrino flux dominates the atmospheric background above 100\,TeV and that the majority of IceCube's detection significance for hard-spectrum sources comes from events above 10\,TeV. 
Fully exploring these higher energies, where fluxes are lower, requires increasing the active effective area of the detector array. The Gen2 design achieves this by roughly doubling the spacing between each string and by deploying more sensitive photon detection modules deeper into the ice, resulting in an increase of 25\% in effective area.
These extensions in the dimensions rooted in the knowledge gained in operating IceCube, which has significantly altered our understanding of the deep glacial ice sheet.
Overall, the initial design of the complete array consists of 120 strings (see Fig.~\ref{ic_detector}), where each string is instrumented with optical sensors with similar sensor spacing to that in IceCube.  

All planned major instrumentation in Gen2 is based on IceCube experience, yet there are differences. A new generation of sensors are being developed for the IceCube Upgrade that have 2-3 times the  photocathode area and that provide additional directional information by employing multiple smaller photomultiplier tubes in each optical module.  
The higher PMT coverage, combined with better understanding of the ice, partially compensates for the substantially increased string spacing.  The electronics and mechanical design principles gleaned from experience with the IceCube DOM (of which greater than 98\% remain fully functional 10 years after completion of IceCube) are carried over, while the gains in extra photocathode area and added directional information of the detected photons result in about 30\% better angular resolution.  As a result, it will be possible to reconstruct horizontal muon neutrino events in Gen2 with an angular resolution approaching $0.1^{\circ}$. We note that the current Gen2 R\&D focuses not just on improving the performance but also on reducing costs and logistic requirements for deploying and maintaining the full-scale OA. This includes developments that reduce the sensor diameter to ease deployment\,\footnote{Based on past drilling simulations and data, an instrument with a diameter reduced by 5\,cm would allow a savings of 15\% in drill time and fuel consumption.}, and an alternative readout scheme that would improve power management and simplify communications. 

\begin{figure}[h!]
    \centering
   \includegraphics[width=0.65\textwidth]{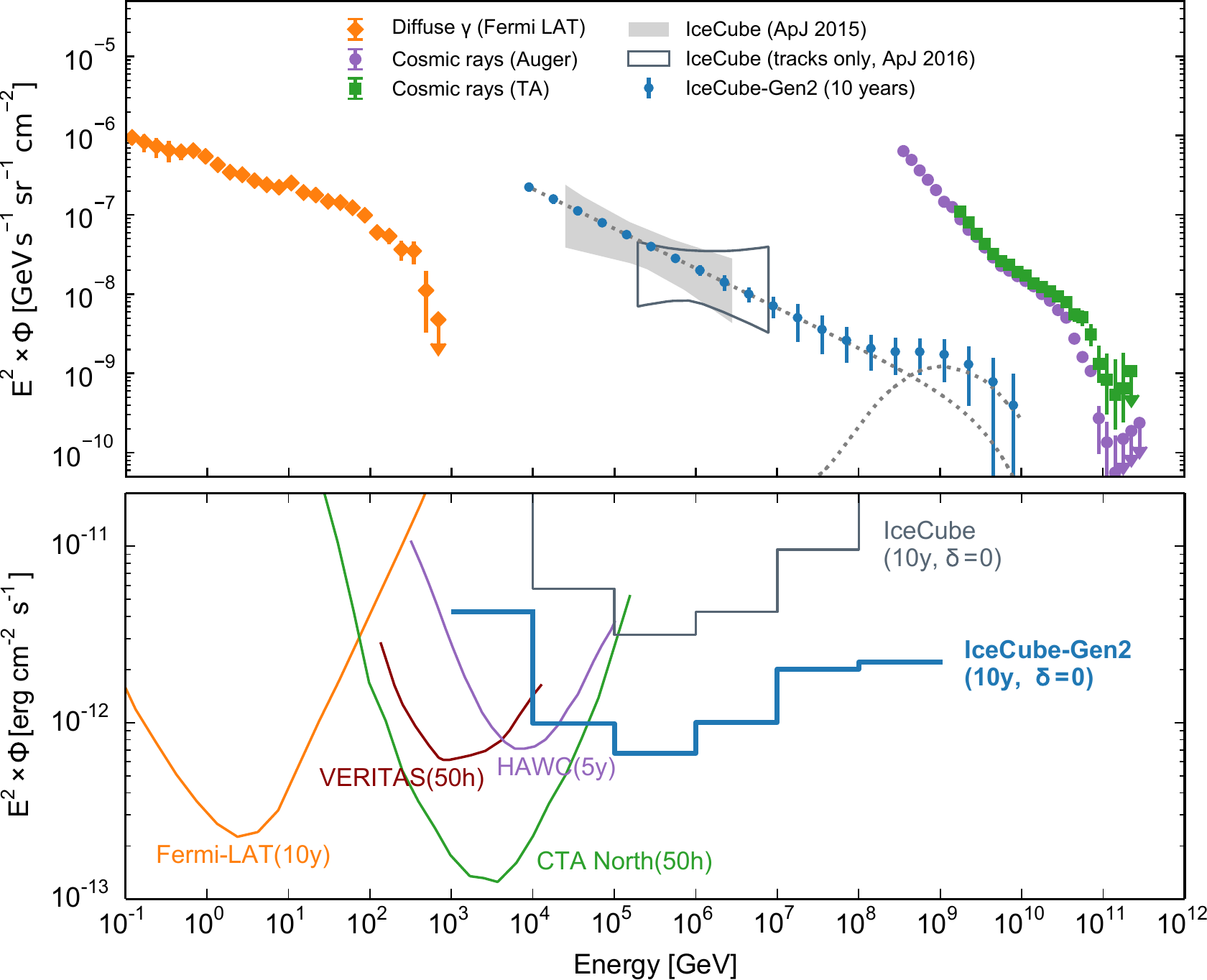}
    \caption{Performance of IceCube-Gen2 in a multimessenger context. Upper panel: The high-energy astrophysical neutrino spectrum, compared to the extragalatic $\gamma$-ray spectrum measured by {\it Fermi}-LAT \cite{Ackermann:2014usa} and the highest energy cosmic ray spectrum measured by Telescope Array~\cite{AbuZayyad:2012ru} and the Pierre Auger Observatory \cite{ThePierreAuger:2013eja}. The grey bands represent the range of neutrino fluxes obtained in~\cite{Aartsen:2015knd,Aartsen:2016xlq}. The blue points are the median flux levels and 68\% confidence intervals that would be obtained from 10 years of IceCube-Gen2 data, assuming that the flux from cosmic neutrino sources continues as $\Phi \propto E^{-2.5}$, and asuming a cosmogenic neutrino flux (10\% proton fraction in the UHECR) as described in \cite{vanVliet:2019nse} (gray dotted lines). Lower panel: Sensitivity of IceCube and IceCube-Gen2 (optical + radio array) to detect the neutrino flux from a point source at the celestial equator with an average significance of 5$\sigma$ after 10 years of observations. It is compared to the sensitivity of selected gamma-ray telescopes to detect gamma-ray point sources after typical observation periods indicated in brackets. The IceCube and IceCube-Gen2 sensitivities are calculated separately for each decade in energy, assuming a differential flux $dN/dE \propto E^{-2}$ in that decade only. For details about the calculation of the sensitivities for the shown gamma-ray telescopes please refer to \cite{Abeysekara:2013tza,lat_performance,veritas_performance,cta_performance} where these sensitivities were taken from. Neutrino fluxes in the upper (lower) panel are shown as the all-flavor (per flavor) sum of neutrino and anti-neutrino flux, assuming an equal flux in all flavors.}
    \label{final_mma_spectrum}
\end{figure}

The IceCube-Gen2 conceptual design studies have evaluated the potential impact on future high-energy astrophysical neutrino measurements, resulting in a baseline configuration shown in Fig.~\ref{ic_detector}. The Gen2 geometry will increase the detection volume from the current 1~km$^3$ by nearly an order of magnitude.  Despite the large gain in sensitivity, the budget required for the OA is comparable to that of IceCube. 
To demonstrate the power of the Gen2 OA conceptual design, a differential flux measurement similar to that observed with IceCube has been simulated. All observed channels were utilized in the study, and the signal hypothesis was a set of E$^{-2.5}$ power law fluxes where each energy segment was permitted to vary independently~\cite{vanSanten:2017chb}. 
The result is that the significant increase in instrumented area of the Gen2 OA, along with improved angular resolution, opens new sensitivity to fluxes from individual  point sources much fainter than the current IceCube limits. 
Figs.~\ref{astro_constraints} and \ref{final_mma_spectrum} show the potential for an approximate factor of five increase in sensitivity to a point source (assuming an E$^{-2}$ flux) made possible with the Gen2 OA. It provides the sensitivity to identify the dominating source population among any of the plausible scenarios.

\begin{figure}
    \centering
    \includegraphics[height=0.4\textwidth]{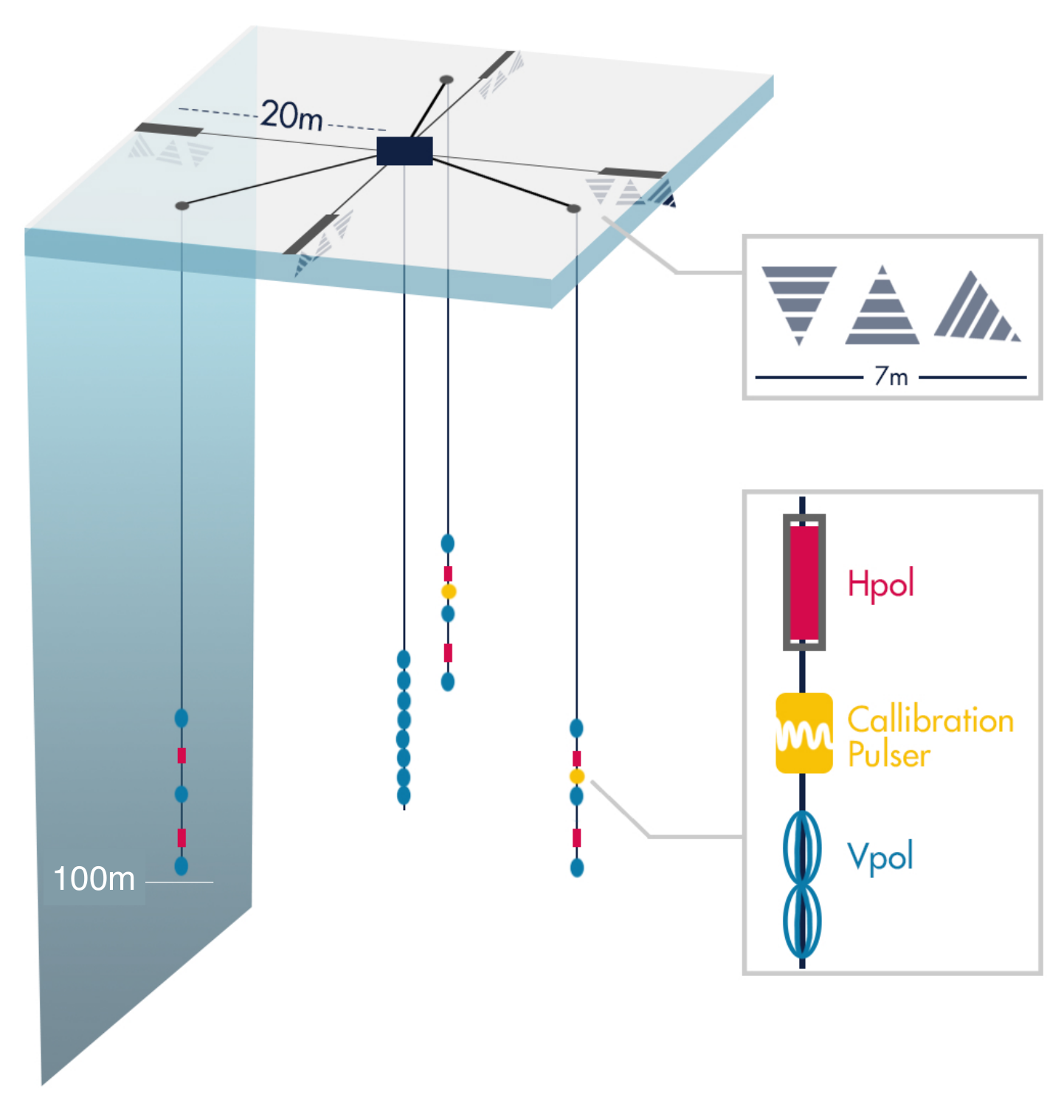}
    \includegraphics[height=0.4\textwidth]{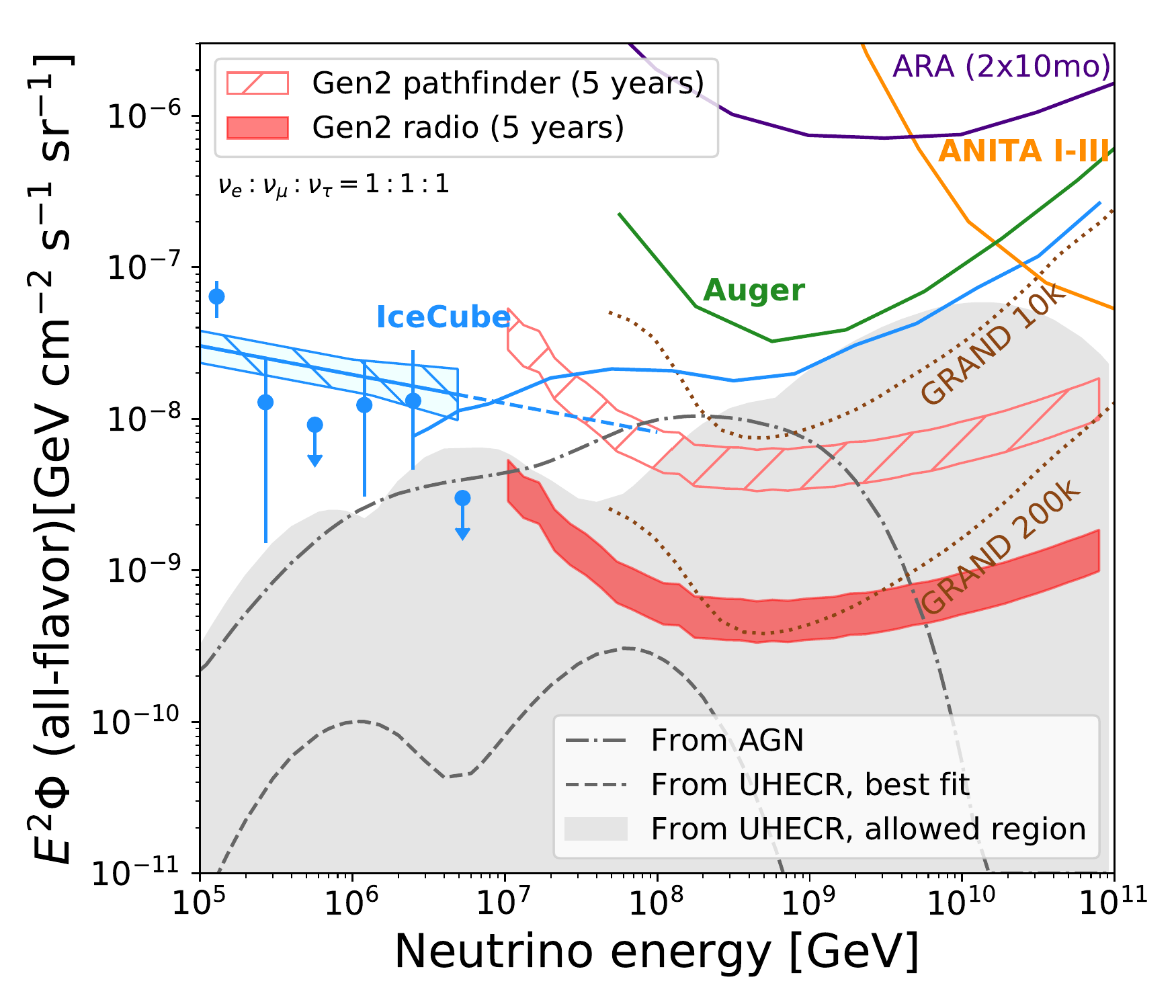} 
      \caption{Left: Possible station design for the radio array of Gen2. Right: The ultra-high-energy neutrino parameter space, including the results from IceCube \cite{Haack:2017dxi,Kopper:2017zzm, Aartsen:2018vtx}, the Pierre Auger Observatory \cite{Aab:2015kma}, ANITA \cite{Allison:2018cxu}, and ARA \cite{Allison:2015eky}, as well as model predictions for AGN \cite{Murase:2014foa} and neutrinos resulting from interaction with photon background \cite{Heinze:2019jou,vanVliet:2019nse}. The sensitivity of 5 years of the Gen2 RA and its planned pathfinder are shown in red bands, assuming a phased array (see below) trigger at 100 meters depth at the South Pole. The width of the band reflects estimated uncertainties. For comparison the sensitivity of two stages of GRAND \cite{Alvarez-Muniz:2018bhp} are shown.
    }
    \label{fig:Radio_component}
\end{figure}

\noindent\textbf{Radio detector array (RA):} 
The Gen2 radio array will provide unprecedented sensitivity at energies above 10 PeV (see Fig.~\ref{fig:Radio_component}), and is essential for probing the ultra-high-energy regime, including a number of key open questions, such as: What is the spectrum of astrophysical neutrinos beyond 10 PeV; What is their flavor composition; and, What is the level of the diffuse cosmogenic neutrino flux from interactions on the photon background? Obtaining the answers to these questions requires an improvement of two orders of magnitude higher exposure than what the current best detectors can achieve at energies above 100 PeV~\cite{2019BAAS...51c.185V}. This suggests a design requirement for future detectors to achieve an approximately equal resolution to a flux characterized by an $E^{-2}$ power law over an energy range from 30 TeV to 30 EeV.  

Radio emission is generated in ice by particle showers through the Askaryan effect \cite{Askaryan1962} and is typically measurable above the thermal noise floor at a shower energy of 10 PeV.
Due to coherence effects, the emission is strong at angles close to the Cherenkov angle, where all emitted radio signals 
arrive coherently. The emitted frequency range is governed by the shower geometry and the spectrum typically shows the strongest contribution between 100 MHz and 1 GHz \cite{AlvarezMuniz:2011ya}. In the time domain, the emission corresponds to a broad-band nanosecond-scale radio pulse, which has been observed both at accelerator experiments \cite{Gorham:2006fy} and in air showers \cite{Aab:2014esa,Schellart:2014oaa,Scholten:2016gmj}. 
The fact that radio emission is generated by purely electromagnetic showers and the electromagnetic component of hadronic showers means that a radio detector is sensitive to all neutrino flavors, albeit with different sensitivities. 

A wide range of experimental efforts has advanced the radio detection technique of neutrinos in the past two decades. They include the ANITA experiment (a balloon-borne radio detector), as well as experimental efforts in Antarctic ice, namely
the Askaryan Radio Array (ARA) at the South Pole \cite{Allison:2011wk, Allison:2015eky} and the ARIANNA detector on the Ross Ice Shelf \cite{Barwick:2014rca, Barwick:2014pca}. The radio attenuation properties of ice have been directly measured at locations in Antarctica and Greenland~\cite{barrella, Barwick:2014rga, avva1}, showing attenuation lengths in excess of 1~km at the South Pole~\cite{Barwick:2005zz}. 
Building on the expertise obtained and data analyzed from all of these experimental efforts, a conceptual design has been developed for the next-generation radio detector in ice.

To achieve the best accuracy in identification and reconstruction of ultra-high-energy neutrinos, antennas with broadband, high gain, and the ability to measure at least two orthogonal polarizations, are needed~\cite{Glaser:2019rxw}. Combining deep with shallow antennas provides a compact station without losing gains in effective volume or reconstruction quantities, and reaps the benefits of a cosmic-ray self-veto \cite{deVries:2015oda,Barwick:2016mxm} and a direct-and-reflected signal \cite{Kelley:2017xjj}. Such a station therefore consists of a cluster of antennas and acts independently in both triggering and reconstructing the shower (see Fig.~\ref{fig:Radio_component}). An array will consist of widely-spaced stations, each monitoring an effectively independent volume. Hence, the total effective volume scales linearly with the number of stations.

Phased-array techniques have the ability to add signals from antennas, thereby digitally emulating a better antenna.  
Since the amplitude of the radio signal scales with energy, a lower detection threshold immediately lowers the energy threshold of the detector and increases the effective volume, making it possible to detect signals of the same strength at larger distances. The principle application of such an interferometric trigger has been developed \cite{Avva:2016ggs} and demonstrated at the South Pole \cite{Allison:2018ynt}.

A Gen2 radio detector station combines shallow antennas (better gain, frequency and polarization sensitivity, as well as an air-shower veto of atmospheric events for improved identification of cosmic neutrinos in the OA), with deep antennas and a phased-array trigger string (more effective volume, better sky coverage). A fraction of events will be detected in all detector components, providing events with the best reconstructed properties. The Gen2 radio array will cover an area of approximately 500\,km$^2$ of the Antarctic ice near the South Pole. 
In the conceptual design 200 stations are envisioned viewing the ice providing unprecedented sensitivity at energies above 10 PeV (see Fig.~\ref{fig:Radio_component}). A  radio pathfinder experiment comprising about 10\% of the final scope is planned  as verification of the concept either at the South Pole or in Greenland.

\noindent\textbf{Surface instrumentation:} The current IceCube observatory has a fully integrated surface array, IceTop, that consists of two ice-Cherenkov tanks above each string of the OA and measures cosmic-ray air showers in the energy range from a few 100\,TeV to about 1\,EeV \cite{IceCube:2012nn}. These measurements have proven valuable to the study of cosmic neutrinos, including the calibration of the in-ice array with muons of air-showers detected by IceTop \cite{Bai:2007zzm}, and for vetoing air-showers in the selection of neutrino candidates \cite{Tosi:2017zho}. 

Looking to the next generation observatory, a desired advance in the accuracy of the cosmic-ray detection in coincidence with the OA and RA may be achieved through a combination of scintillator, radio and air Cherenkov detector elements to achieve complementary sensitivity to the electromagnetic and muonic shower components in a hybrid surface array~\cite{Holt:2019fnj}. 
Such a surface detector system also allows for cross-calibration of IceCube's absolute energy scale with those of other air-shower detectors \cite{Apel:2016gws, GSF_ICRC:2017}. 
The anticipated boost in measurement accuracy and the increase in sky coverage of the hybrid design are expected to provide systematics limited results for studies of the most energetic Galactic cosmic-rays before the end of the decade \cite{Schroder:2018dvb, Haungs:2019ylq}. As a part of IceCube-Gen2, an increase in the geometrical size of the surface array over the current IceTop (by approximately an order of magnitude) is anticipated, thus overcoming this foreseen limitation while also significantly enhancing the ability to veto atmospheric neutrinos and muon tracks in the selection of high-energy astrophysical neutrino candidates from the Southern Hemisphere. The latter is particularly important for studying the astrophysical neutrino flux at lower energies ({\it i.e.}, at and below 10 TeV), as seen in Fig.\,6.

\noindent\textbf{A wide-band neutrino facility:} 
The South Pole provides a unique logistical hub for a wide-band neutrino facility spanning an energy range from 10 TeV to beyond 100 EeV.  The existing IceCube Observatory and the IceCube Upgrade now under construction will be fully integrated in the ultimate Gen2 facility, providing energy coverage that will range from 1 GeV to 100 EeV.  In this way, IceCube-Gen2 will be the world's preeminent atmospheric and astrophysical neutrino observatory.  
Such a facility provides access to world-class science, including many topics not explicitly covered in this White Paper, such as: indirect searches for dark matter, particle physics at the highest energies ({\it e.g.} measurements of cross-section and inelasticity, tests of Lorentz violation, etc.), and the unique capability to detect MeV-scale supernova neutrinos from well beyond our galaxy.\footnote{A supernova core collapse in our galaxy would result in a neutrino burst whose time profile could be recorded with high statistics and ms resolution.}

IceCube-Gen2 will leverage many of the resources and much of the expertise of the current IceCube and IceCube Upgrade arrays. Gen2 will utilize the well-developed operational infrastructure and experience of the existing detector, including the counting house, the data handling systems, and data management in the northern hemisphere. Success within the South Pole environment is made possible only via the NSF's Amundsen-Scott South Pole station, a premier scientific facility that provides key resources year-round to the projects, including power, communications and logistical support for operations. To mitigate impact on those established resources, it is anticipated that the additional logistical support, including shipping and personnel, for the construction period of IceCube-Gen2 would be requested - based on IceCube and Upgrade construction as a key point of reference -- within the scope of the project's funding.

\pagebreak
\clearpage
\subsection*{IceCube-Gen2 in the Era of Multimessenger Astrophysics}

\begin{wrapfigure}{r}{9cm}
\centering
\vspace{-20pt}
\includegraphics[width=\linewidth]{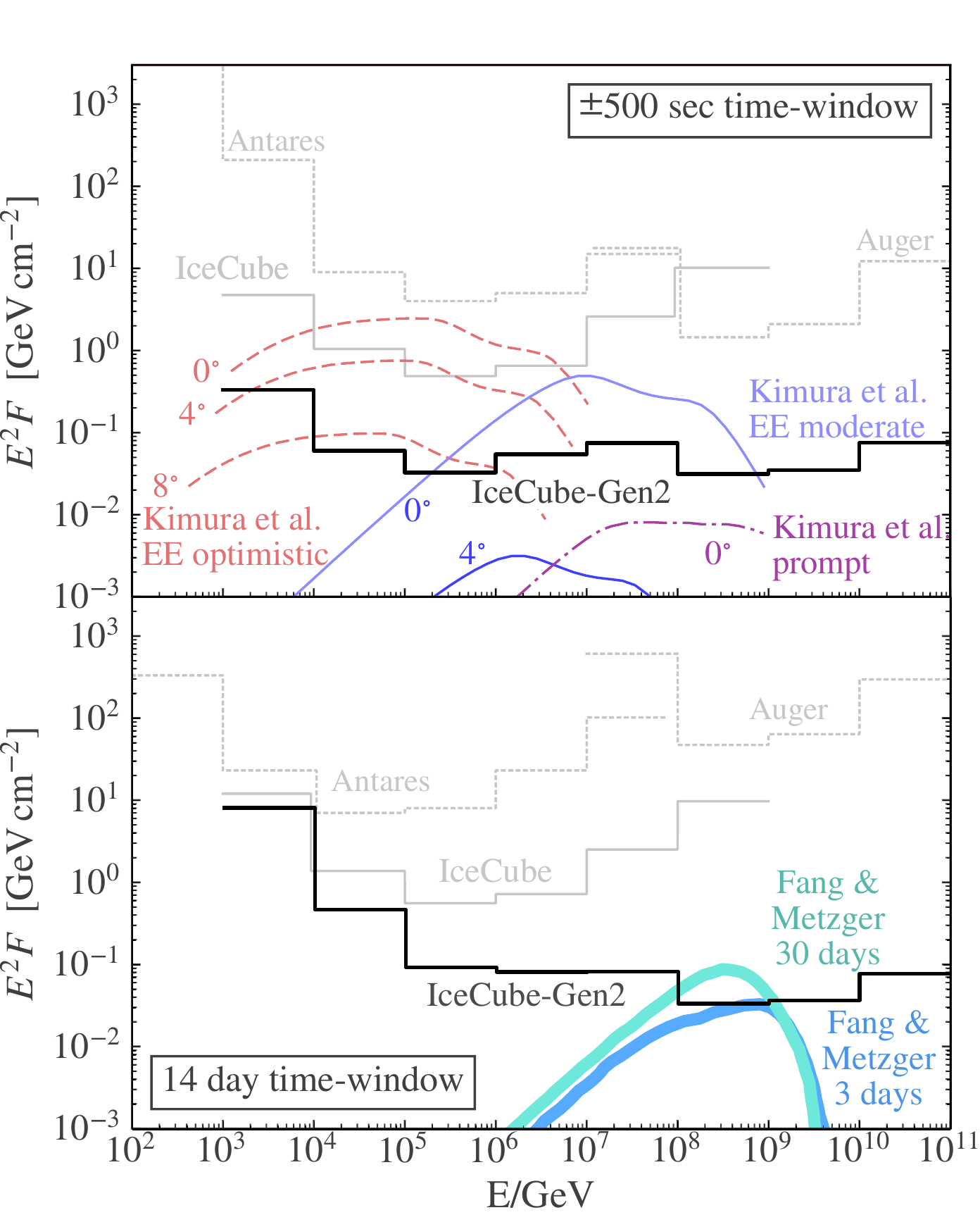}
\caption{Upper limits (at 90\% CL) from various instruments on the neutrino spectral fluence from GW170817 during a $\pm$500~s window centered on the gravitational wave (GW) trigger time (top panel), and a 14-day window following the trigger (bottom panel). For comparison, the sensitivity of IceCube-Gen2 (at 90\% CL) to an event at a similar position on the sky (solid black line) 
is presented. Also shown are several predictions by neutrino emission models \cite{2017ApJ...849..153F,2018PhRvD..98d3020K} scaled to a distance of 40 Mpc. Separate curves are displayed for different components of the emission (prompt and extended (EE)), observation angles relative to the jet axis, and time scales of the emission. See \cite{2017ApJ...850L..35A} for details. Limits and sensitivities are calculated separately for each energy decade, assuming a spectral fluence $F(E) \propto E^{-2}$ in that decade only. All fluences are shown as the per flavor sum of neutrino and anti-neutrino fluence, assuming equal fluence in all flavors, as expected for standard neutrino oscillation parameters. Figure adopted from \cite{2017ApJ...850L..35A}.}
\label{fig:gen2_nsmerger_sens}
\vspace{-20pt}
\end{wrapfigure}

Multimessenger astrophysics is now a reality. Recent advances and discoveries with gravitational waves and high-energy neutrinos have opened new long-anticipated windows on the universe, and have begun to transform our understanding of the high-energy astrophysical universe and the mechanisms at work in the most energetic phenomena in the cosmos. One of the most important messages emerging from the IceCube measurements to date is that for the high-energy cosmic neutrino flux a prominent and surprisingly important role is played by protons relative to electrons in the extreme non-thermal universe.  The matching energy densities of the extra-galactic gamma-ray flux detected by $Fermi$-LAT, the high-energy neutrino flux measured by IceCube, the cosmic ray flux at the highest energies, suggest that rather than detecting some exotic sources, IceCube is now providing a new path to study their common origin, the most extreme accelerators in the Universe (see Fig.\,\ref{final_mma_spectrum}). Exploring these via the information provided by different messengers is expected to unlock new knowledge of the underlying astrophysical processes, including those opaque to electromagnetic radiation.  

IceCube-Gen2 has a design sensitivity that will deliver the answers to many of the current most pressing questions in the field, including resolving the sources of the large flux of cosmic neutrinos (and hence the sources of cosmic rays).
IceCube-Gen2 will significantly impact complementary measurements between neutrinos and high-energy gamma-rays (see Fig.\,\ref{final_mma_spectrum}), but also connect to the highest energy cosmic ray sources via its significantly expanded sensitivity towards higher energies. In addition, with the continued advancement of gravitational wave measurements, synoptic optical surveys such as LSST, X-ray and radio surveys, the reach of IceCube-Gen2 to probe a number of key models of extreme cosmic events, including those associated with binary neutron star mergers (see Fig.\,\ref{fig:gen2_nsmerger_sens}), will finally be within reach in the next decade. 

\pagebreak
\clearpage

\bibliographystyle{hieeetr}
\section*{\refname}
\bibstyle{plain}
\bibliography{references}

\end{document}